# Storing a single photon as a spin wave entangled with a flying photon in telecomband

Wei Zhang, Dong-Sheng Ding[†], Shuai Shi, Yan Li, Zhi-Yuan Zhou, Bao-Sen Shi[*], Guang-Can Guo

[1]*Key Laboratory of Quantum Information, CAS, University of Science and Technology of China, Hefei, Anhui, 230026, China*

[2]*Synergetic Innovation Center of Quantum Information & Quantum Physics, University of Science and Technology of China, Hefei, Anhui, 230026, China*

Corresponding author: [†]*dds@ustc.edu.cn*

[*]*drshi@ustc.edu.cn*

Quantum memory is an essential building block for quantum communication and scalable linear quantum computation. Storing two-color entangled photons, with one photon being at telecom-wavelength while the other photon being compatible of quantum memory, has great advantages toward the realization of the fiber-based long-distance quantum communication with the aid of quantum repeaters. Here, we report an experimental realization of storing a photon entangled with a telecom photon in polarization as an atomic spin wave in a cold atomic ensemble, thus establishing the entanglement between the telecom-band photon and the atomic-ensemble memory in polarization degree of freedom. The reconstructed density matrix and the violation of Clauser–Horne–Shimony–Holt inequality clearly show the preservation of quantum entanglement during storage. Our result is very promising for establishing a long-distance quantum network based on cold atomic-ensembles.

To realize a long-distance quantum communication, a quantum repeater has to be used to overcome the problem of communication fidelity decreasing exponentially with the channel length [1-3]. Quantum memories for light [4, 5], which have been realized successfully during the past decade in many systems including a cold [6-11]/hot atomic system [12-15], a solid matter [16-18], a diamond [19], and others [20-22], are key components consisting of a quantum repeater. The realization of quantum repeaters requires the storage of quantum entanglement at local nodes and swapping the entanglement between adjacent nodes, which can be achieved by combining the two photons from different nodes intermediately [23-27]. Therefore, the photon travelling from the node to the location where Bell state measurement is performed in the low-loss window of optical fiber is preferred. By this way, the number of repeaters could be reduced significantly. However this is very hard to realize in atomic media due to the lack of the accessible energy levels in the scheme of Duan-Lukin-Cirac-Zoller [1]. This problem can be overcome using two different ways: one is the frequency conversion in atomic ensembles [28, 29] or in nonlinear crystals [30], in which three or two-order nonlinear process has to be used. Another solution is to avoid this problem from the start by interfacing sources of entangled photons, where one photon of each pair is at a telecommunication wavelength, with the other photon being compatible of an optical quantum memory [25, 17, 18], as pointed out in Ref. 5. Kuzmich's group firstly established a quantum memory with telecom-wavelength conversion in 2010 [29] and later succeeded making it compatible of entanglement [28] in the DLCZ scheme using frequency conversion method. However, an experimental realization of generating two-color entanglement in one atomic ensemble and storing this entangled photon in another ensemble as an atomic spin has never been reported before.

Here, we report the first experimental storage of a two-color polarized entanglement in a cold atomic ensemble using the electromagnetically induced transparency (EIT) protocol, by which the entanglement between the atomic spin wave and the photon in telecomband is established. In our experiment, the polarization-entangled photons with the wavelength of one photon matching the transition wavelength of rubidium (Rb) atom while the other being in telecomband are generated directly by spontaneously cascaded emission in one cold atomic ensemble, then this two-photon state is improved to be maximally entangled by letting one photon passes through a phase-insensitive Mach-Zehnder interferometer with an attenuation plate inside. After that, this photon is stored in another cold atomic ensemble embedded in the interferometer. In this way, the entanglement is established between the atomic ensemble and the telecomband photon which transmits in an optical fiber. After 100-ns storage, we convert the atom-photon entanglement to the photon-photon entanglement and check their entanglement. We reconstruct the density matrix for the photon-photon entanglement with a fidelity of 88.8±4.4%, and obtain the violation of Clauser–Horne–Shimony–Holt (CHSH) inequality by 3.2 standard deviation without any noise correction. All results clearly show the preservation of the entanglement during the storage. Our work proves a successful quantum memory for two-color polarized entanglement in a cold atomic ensemble that constitutes a basis for building up a fiber-based long-distance quantum network.

Before showing our main results, we want to mention a fact that an erbium-crystal or an optical crystal fiber can store a photon in the telecomband because these materials have suitable transition energy levels, however storing non-classical light with a fidelity higher than classical limit is a hard task in these materials due to improper relaxation dynamics for pumping [31] or noise issues [32]. Recently, Tittel's group demonstrated the storage and recall of an entangled 1532-nm-wavelength photon in an ensemble of cryogenically cooled erbium ions doped into a 20 m-long silica fiber, using a photon-echo quantum-memory protocol [22].

The medium used here to generate a two-color polarization entangled photon pair is an optically thick ensemble of $^{85}$Rb atoms trapped in a two-dimensional magneto-optical trap (MOT) [33]. The experiment setup is presented in Fig. 1. A pair of non-maximally entangled photons with 795-nm ('Signal 2') and 1475-nm ('Signal 1') wavelengths respectively is created from the cold atomic ensemble in MOT A. To achieve this, 780-nm ('Pump 1') and 1530-nm ('Pump 2') lasers with orthogonal polarization collinearly pump the atomic ensemble under the condition of near two-photon resonance. Using a series of mirrors and lenses, the Signal-2 photon is delivered to the MOT B for subsequent storage, while the Signal-1 photon is coupled into a 200-m long optical fiber. A self-locked Mach-Zehnder interferometer with two half-wave plates (HWPs), two quarter-wave plate (QWPs) and an attenuation plate used here has two important effects: one is to improve the entanglement of the photon pair, the other is to guarantee the same memory efficiency for the differently polarized single-photon state, see Appendix B.

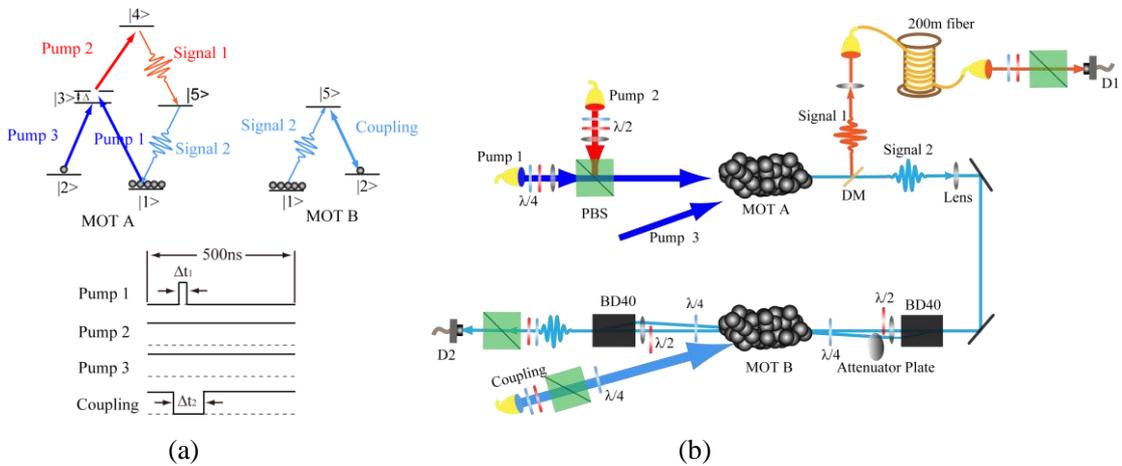

Fig 1. (a) Simplified energy diagram and timing sequence for the generation of polarization

entanglement. Pump 1 is a pulse of duration $\Delta t_1$=20 ns modulated by the acousto-optic modulator (AOM). $\Delta t_2$ represents the storage time. The single-photon detuning $\Delta$ is +130 MHz. $|1>=|5S_{1/2},F=2>$, $|2>=|5S_{1/2},F=3>$, $|3>=|5P_{3/2},F=3>$, $|4>=|4D_{3/2},F=2>$, $|5>=|5P_{1/2},F=3>$. (b) Simplified experimental setup. PBS: polarization beam splitter; $\lambda/2$ half-wave plate; $\lambda/4$ quarter-wave plate; $D_1$: a free running In-GaAs single-photon detector, (ID Quantique ID220-FR-SMF); D2: avalanche diode (Perkin-Elmer SPCM-AQR-15-FC); BD40:it can separate input light into two orthogonally polarized beams with 4.0 mm beam separation (Thorlabs BD40).The Coupling light is above the two beams passing through BD40 and have 1.5 ° with both of them. The power of Pump 1, Pump 2, and the coupling light are 0.1 mW, 8 mW, and 20 mW, respectively.

Our system works at the repetition rate of 100 Hz. A period T=10 ms includes the MOT trapping time (including initial state preparation) of $t_{MOT}$=8.7 ms, and the operation time of $t_{duty}$=1.3ms, which contains 2600 cycles with a cycle time of 500 ns. During each cycle, Pump-2 and Pump-3 light beams are kept open, the Pump-3 at 780-nm wavelength is used to repump atoms in the MOT A to the initial state |1>. Pump-1 light beam is shaped to be a pulse with 20-ns full width at half maximum (FWHM) every 500 ns. Both Pump-1 and Pump-2 light beam are focused on the atomic ensemble in MOT A with a focal length of 500 mm. Also, during the storage, Signal-2 photons are focused on the atomic ensemble in MOT B. The optical depth (OD) of atomic ensemble in MOT A and MOT B is 20, 50 respectively.

Here the source are generated based on spontaneously four-wave mixing in atomic ensembles, in which single atom are pumped to state |4> through two-photon resonance, then back to state |1> with the emission of Signal 1 and Signal 2. The spontaneously four-wave mixing are widely used in diamond-type [34, 35], ladder-type [36, 37] and Lambda-type [6] configuration in atomic media.

In this experiment, the non-maximally two-color entangled photons, generated through spontaneously cascaded emission in a diamond-type configuration using orthogonal polarized pump light beams, can be expressed as a two-photon state [38]:

$$|\psi_1>=\cos\eta_f |H_{S_1}V_{S_2}>+e^{i\phi_f}\sin\eta_f |V_{S_1}H_{S_2}> \qquad (1)$$

where $|H_{S_1}>(|H_{S_2}>)$ and $|V_{S_1}>(|V_{S_2}>)$ represent, respectively, the horizontal and vertical polarizations of Signal 1 (Signal 2). $\phi_f$, a controllable parameter, represents the phase shifts induced by the pump light in the atomic media and the various optical elements. The mixing angle $\eta_f$, determined by the dipole matrix elements for different polarization, is very sensitive to the two-photon detuning of pump light, see Appendix A. In the experimental process, we find $\eta_f \approx 1.25\,\pi/4$, and $\phi_f \ll 1$ at -20MHz of two-photon detuning. The photon pair is improved to be maximally entangled after the Signal-2 photon passes through a self-locked Mach-Zehnder interferometer with an attenuation plate inside, which is used to slightly attenuate the horizontal Signal-2 photon to balance two terms in the two-photon state. In this case, the two-photon state can be written as:

$$|\psi_{input}> = \frac{1}{\sqrt{2}}(|H_{S_1}V_{S_2}> + |V_{S_1}H_{S_2}>) \qquad (2)$$

By this skill, the maximally entangled state was achieved between the Signal-2 photon and the Signal-1 photon transmitting in a 200-m long optical fiber. Then we opened the MOT B and stored the Signal-2 photon. Thus the entanglement was established between the collective atomic excitation (also called spin wave) and the Signal-1 photon, which can be denoted as,

$$|\psi_{aS_1}> = \frac{1}{\sqrt{2}}(|H_{S_1}V_a> + |V_{S_1}H_a>) \qquad (3)$$

where $|H_a>(|V_a>)$ represents the collective atomic excitation.

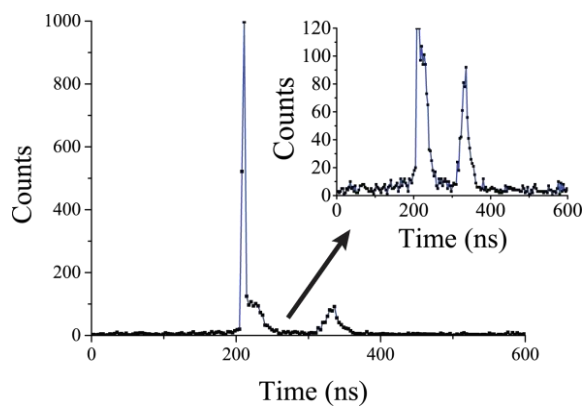

(a)

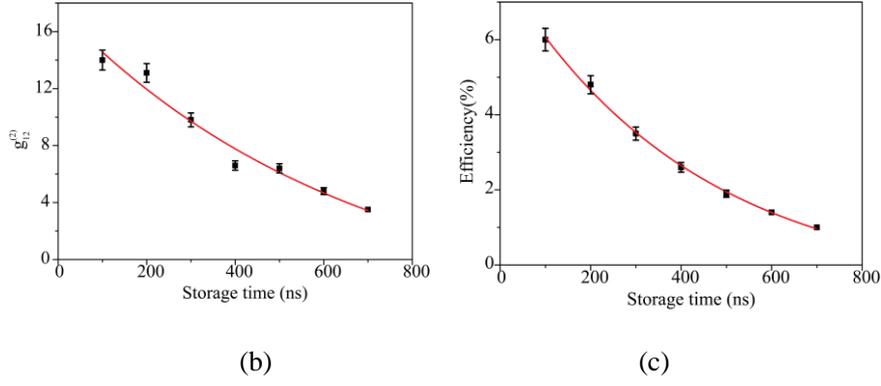

Fig. 2. Performance of the quantum memory. (a) Cross-correlation between Signal 1 and Signal 2 photons after 100-ns storage. (b), cross-correlation $g^{(2)}_{12}$ versus storage time. (c), storage efficiency versus storage time.

Fig. 2 (a) shows the cross-correlation signal after 100-ns storage. And Fig. 2 (b) and (c) described $g^{(2)}_{12}$ and storage efficiency decayed with the storage time. Experimentally, after a 100-ns storage, the collective atomic excitation in MOT B is read out to be a single photon. During this process, the storage time in MOT B should be shorter than the time delay of Signal 1 in optical fiber in order to guarantee the storage of the entanglement. In our experiment, the time delay of the Signal-1 photon, determined by the transmission time in the optical fiber, is about 1μs. The storage efficiency for the programmed time of $\Delta t_2 = 100 ns$ is ~ 6%. To demonstrate experimentally whether the entanglement is preserved or not, individual projection measurements related to projecting the two-photon state into the four basis, $|H>, |V>, (|H>+|V>)/\sqrt{2}$, $(|H>-i|V>)/\sqrt{2}$ are performed on each photon before and after storing the 795-nm photon. According to the measurement results, we reconstructed the density matrix of the photon pair before and after storage [39], which is shown in Fig. 3.

From the reconstructed density matrices, we calculate the fidelity which quantifies how closely the one state resembles the others. As is well known, the ideal retrieved polarized entangled state should be,

$$|\psi_{ideal}> = \frac{1}{\sqrt{2}}(|H_{S_1}V_{S_2'}> + |V_{S_1}H_{S_2'}>) \qquad (4)$$

where $|H_{S_2'}>$ denotes the photon retrieved from the atomic spin wave. The fidelity of the photon pair state before storage compared with the ideal state is 88.1±2.6% obtained by comparing the two-photon state density matrix $\rho_{input}$ with the ideal density matrix $\rho_{ideal}$, using the formula $F_1 = Tr(\sqrt{\sqrt{\rho_{input}}\rho_{ideal}\sqrt{\rho_{input}}})^2$. Figure 3 depicts the real and imaginary parts of the density matrix for the state before and after storage. The storage fidelity $F_2 = Tr(\sqrt{\sqrt{\rho_{output}}\rho_{input}\sqrt{\rho_{output}}})^2$, which quantifies how closely the output state resembles the input state, is 88.8±4.4%.

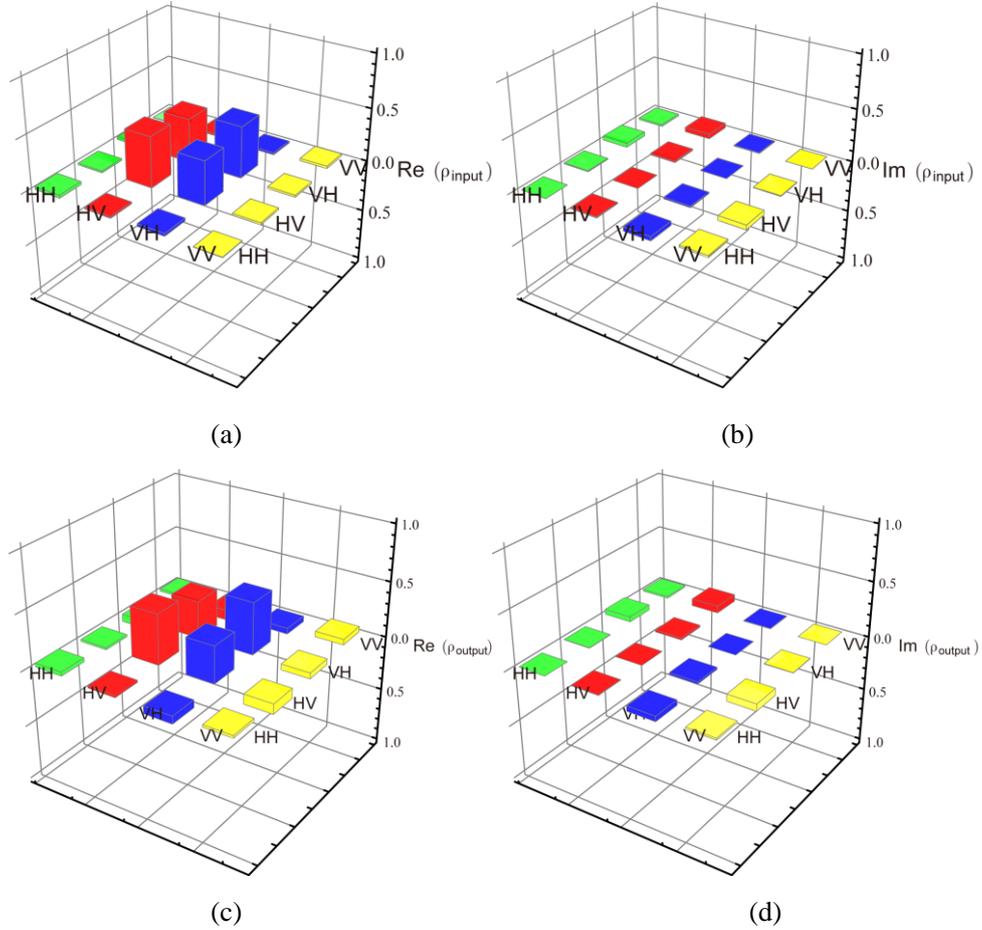

Fig. 3 Real (a, c) and imaginary (b, d) parts of the reconstructed density matrix of the two-photon state before (a, b) and after (c, d) storage, respectively, all data are raw without any noise correction.

Furthermore, we demonstrate the entanglement after storage through checking the violation of the CHSH-type Bell inequality. We measure the correlation function $E(\theta_1, \theta_2)$, which can be calculated from each integration of the coincidence, with $\theta_1(\theta_2)$ being the polarization angles of half-wave plate for the Signal-1 (Signal-2) photon. We obtain the CHSH parameter $S = |E(\theta_1,\theta_2) - E(\theta_1,\theta_2') + E(\theta_1',\theta_2) + E(\theta_1',\theta_2')|$ with $\theta_1=0$, $\theta_2=\pi/8$, $\theta_1'=\pi/4$ and $\theta_2'=3\pi/8$. Here, the S values obtained are 2.49±0.06 before storage and 2.38±0.12 after 100-ns of storage without any noise correction.

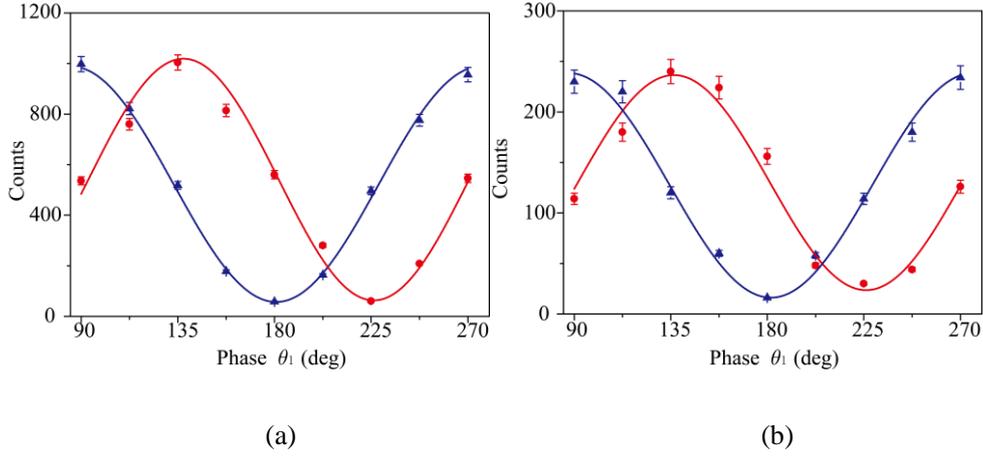

(a) (b)

Fig. 4 Two-photon interference (a) before storage and (b) after storage. The blue (red) curve represents the coincidence rate with the Signal-1 photons projected onto state $|H>$ ($(|H> - |V>)/\sqrt{2}$). Error bars are ±1 standard deviation.

We also study the two-photon interference in Fig. 4. We measure the coincidence counts while the Signal 1 photon is in state $|H>$ ($(|H> - |V>)/\sqrt{2}$) basis for varying polarization angles $\theta_2$ of the half polarization plate for Signal 2 before and after storage. The results are given in Fig. 3. The fitted data showed the visibilities are 88.3%±2.7% before and 81.2%±4.0% after storage. Both values are larger than the threshold of 70.7%, and hence provide clear evidence of non-classical interference, showing that entanglement is preserved.

In this experiment, the overall efficiency of storage is lower than 6% as depicted in Fig. 2 (c). This is mainly due to the spectrum mismatch of Signal-2 photon with our memory system. The Signal-2 single photon has a broad spectrum band (100~200MHz) which almost centres on the transition of $5S_{1/2}(F=2) \to 5P_{1/2}(F=3)$, measured by scanning a Fabry-Perot (FP) cavity through temperature control. Usually, a broad bandwidth wavepacket can be stored in the cold atomic system through Raman memory protocol, which requires a sufficient spectral separation from resonant transitions [10]. However, in our experiment, this broadband single-photon wave packet of Signal 2 is almost on resonance, which makes both EIT protocol and Raman protocol difficult to be effectively utilized. Finally, we take an improved-EIT protocol with carefully parameter adjustment and realize the storage with limited efficiency, see Appendix B.

Another important aspect is the noise issue. Due to the low memory efficiency of this process, the scattering noise coming from coupling laser has to be carefully eliminated. We use three home-made Fabry-Perot cavities before detector D2 with $10^7$:1 ratio and 40% transmission totally and a nearly backward coupling to reduce the scattering noise. The noise from Pump 1 (780nm) is totally blocked by four 808 nm clean-up filters (Semrock LL01-808) with a 22° inclination. For 1475nm single photons, we use three fibre-gratings and a dichroic mirror (which are not depicted in Fig.1) to avoid the Raman scattering generated by the coupling of strong Pump 2 (1530nm) into 200-m long fibre.

In general, memory time can be improved by compensating the magnetic field or by using magnetic field-insensitive states and reducing atomic motion by using optical lattice, a millisecond even hundred millisecond storage time could be achieved [40, 41, 29]. In addition, the dynamic decoupling method can also be used to improve the storage time [42]. Detection is mainly limited by the efficiency of the detector at the telecom-wavelength, which is 10% with a 1-μs dead time here; hence, this experiment can be improved significantly if superconducting detectors can be used. A high-finesse cavity (~10MHz) can be used to improve the memory process [9], and thereby high memory efficiency can be obtained.

In summary, we have experimentally realized the preparation and storage of the two-color photonic polarization entanglement, in which a 795-nm photon is stored in an atomic ensemble while a telecom-wavelength (1475-nm) photon transmits to a distant node by a 200-m long fibre. Our work shows a basic memory element for future fibre-based long-distance quantum communication.

**Acknowledgements**
This work was supported by the National Fundamental Research Program of China (Grant No. 2 011CBA00200), the National Natural Science Foundation of China (Grant Nos. 11174271,

# APPENDIX A: DETAILS ABOUT SOURCE

**1. Coupling efficiency of Signal 1 and Signal 2 photons in details**

The coupling efficiencies for two paths of Signal 2 are 75%. Signal 1 photons are collected through twice coupling: firstly from free space to fibre, then from free space to detector 1, with the coupling efficiency of 82% and 80% respectively. Before collected into a 200-m long fibre, Signal 1 is filtered using a dichroic mirror (30 dB with 99% transmission,) and three fibre-gratings (totally 90 dB with 95% transmission). By this way, the Raman scattering noise caused by Pump 2 coupled in the 200-m fibre is totally eliminated (here $XdB = 10Log_{10}(\frac{Signal1Transmission}{Pump2Transmission})$). Over all, the coupling efficiency of Signal 1 is around 60%.

**2. Characterizing the non-classical correlation between Signal 1 and Signal 2 photons**

Figure 5 shows the coincidence counts between Signal 1 and Signal 2 before storage. The Signal 1 and Signal 2 photons are non-classical correlated in time domain, which could be proved by checking whether the Cauchy-Schwarz inequality is violated or not. It is well known that classical light satisfy the inequality of $R = \frac{[g_{s1,s2}(\tau)]^2}{g_{s1,s1}(0)g_{s2,s2}(0)} \leq 1$. If $R>1$, light is non-classical. Where $g_{s1,s2}(\tau)$, $g_{s1,s1}(0)$, and $g_{s2,s2}(0)$ are the normalised second-order cross-correlation and auto-correlation of the photons respectively. Before storage, we measured $g_{s1,s1}(0)=1.2\pm0.1$, $g_{s2,s2}(0)=1.38\pm0.12$, $g_{s1,s2}(\tau) =150\pm7$ at $\tau=202$ ns. After 100-ns storage, we got $g_{s1,s1}(0)=1.2\pm0.1$, $g_{s2,s2}(0)=2$ (Since the measured $g_{s2,s2}(0)$ after storage included auto-correlation of the leaked signal and auto-correlation of retrieved signal, so the measured data did not represent the real auto-correlation of stored signal. We took the fact that the retrieved photons exhibit photon statistics typical of thermal light), $g_{s1,s2}(\tau) =14\pm0.7$ at $\tau=325$ ns. So we got R=13587±3100 before storage and R=82±14 after 100 ns storage. Both were much larger than 1, the Cauchy-Schwarz inequality was strongly violated, clearly demonstrating the non-classical correlation existed between signal 1 and signal 2 photons before and after storage.

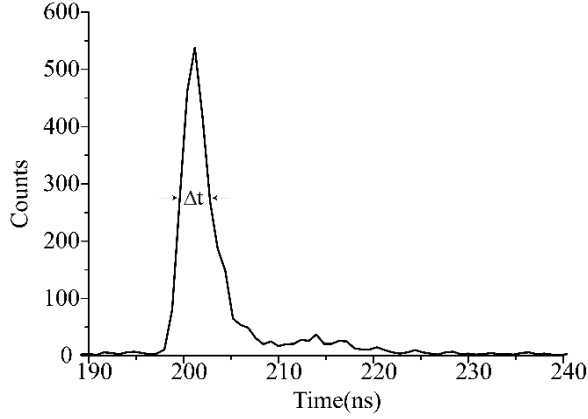

Fig. 5. Cross-correlation between Signal 1 and Signal 2 photons without storage.

## 3. Checking the single-photon properties of the Signal 2 photon

We also characterized the single photon property of the Signal 2 photon by checking a heralded auto-correlation parameter $\alpha$ ($\alpha = \dfrac{P_1 P_{123}}{P_{12} P_{13}}$, where $P_1$ was the signal 1 photon counts, $P_{12}$ and $P_{13}$ are the two-fold coincidence counts between the signal 1 photon and the two separated signal 2 photon by a beam splitter, $P_{123}$ is the three-fold coincidence counts), which were of 0.04±0.02 before storage and 0.3±0.1 after 100-ns storage. $\alpha$ <0.5 for Signal 2 before and after storage clearly demonstrated the single-photon nature.

## 4. Preparing two-color entangled photon pair

Experimentally, we generated two-color entangled photon pair via spontaneously cascaded emission in the cold atomic ensemble trapped in MOT A

$$|\psi_1\rangle = \cos\eta_f |H_{S_1} V_{S_2}\rangle + e^{i\phi_f} \sin\eta_f |V_{S_1} H_{S_2}\rangle \qquad (5)$$

We found that the mixing angle $\eta_f$ was determined by the dipole matrix elements for different polarization, was very sensitive to the two-photon detuning of pump lights. From the expression above we know $\tan^2 \eta_f$ represents the relative intensity of $V_{S_1} H_{S_2}$ to $H_{S_1} V_{S_2}$. We experimentally observed that $\tan^2 \eta_f$ was dependent on the two-photon detuning of pump light as depicted in Fig. 6.

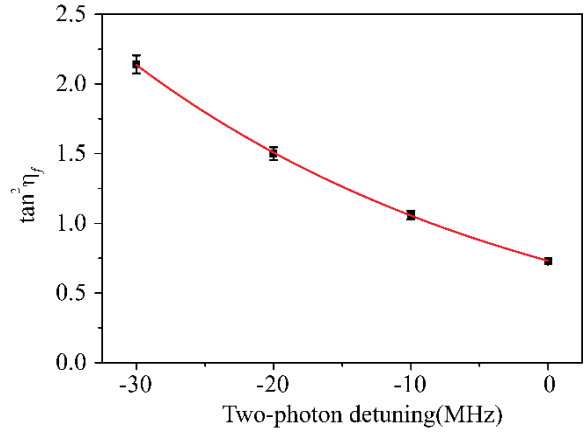

Fig. 6 $\tan^2 \eta_f$ versus two-photon detuning

At same time, with different two-photon detuning of pump light, the time distribution of the generated signal 2 (795nm) was different. Fig. 7 illustrated the experimental results, where the time distribution of the Signal 2 is obtained by a trigger which is synchronous with the Pump 1 pulse. The physical explanations of this phenomenon are not clear at present, which needs further investigation.

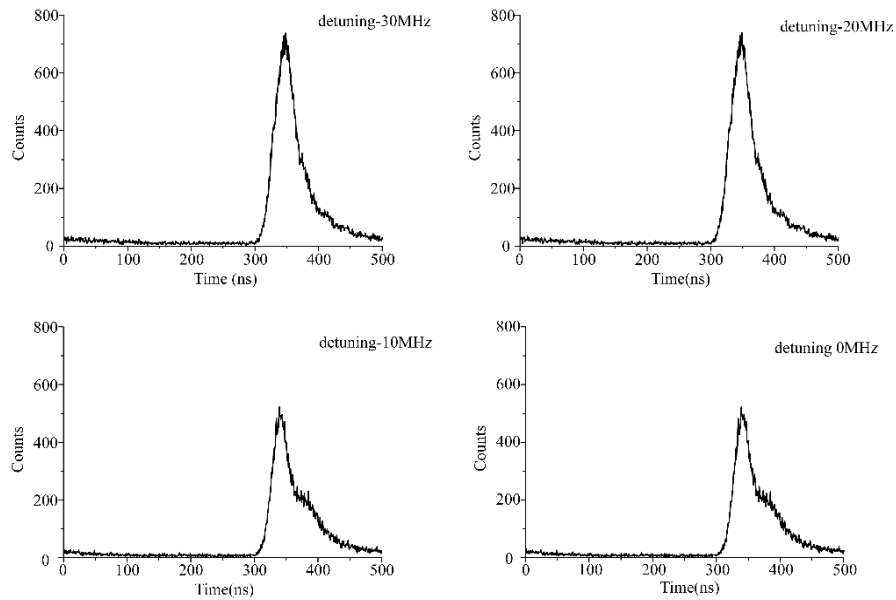

Fig. 7 Time distribution of Signal 2 versus two-photon detuning

We can see it from Fig. 7 that a small two-photon detuning of two pump light makes the time distribution of generated Signal 2 photon be narrower. From Fig. 5, we know the FWHM of two-photon wave-packet is $\Delta t \approx 7ns$, so the real single photon wave-packet of Signal 2 is even narrower than 7 ns, so a narrower time distribution of Signal 2 photon is preferred for efficient storage. Also considering the imbalance caused by two-photon detuning ($\tan^2 \eta_f$ relates to two-photon detuning, and $\tan^2 \eta_f = 1$ is the ideal parameter), we chose the detuning of -20MHz as our experimental parameter. Under this condition, $\tan^2 \eta_f \sim 1.5$ and the FWHM of time distribution of Signal 1 is about 50 ns.

## APPEXDIX B: MEMORY METHOD

**1. Characterizing memory protocol: EIT protocol**

We used a cavity (FWHM=200MHz) to determine the centre frequency of Signal-2 photons, and found it was near to the resonance of the atomic transition |1>->|5> (|$5S_{1/2}$F=2>->|$5P_{1/2}$F=3>). Then we used an improved-EIT protocol and successfully achieved the storage of this broad bandwidth and on-resonant single photon. Fig. S4 illustrated the transmission spectra of a coherent 50-μs probe light (Black) versus probe detuning from the transition of |1>->|5> under the condition of absence (Blue) and presence (Red) of the coupling light with a power of 20 mW in ensemble of MOT B. Here the probe light and coupling light had $\sigma^+$ and $\sigma^-$ polarization. We wanted to emphasize that this $\sigma^+$-$\sigma^-$ polarized (backward) configuration had broader and higher transmission of EIT window than that with usual $H-V$ polarized configuration. Here, we adopt nearly backward operation of probe-coupling (178.5 °angle between them) light to minimize the scatter noise. In addition, three home-made cavities were used to further filter the scattering photons. The optical depth (OD) in MOT B was estimated to be ~50. We observed a broad EIT window of 20 MHz.

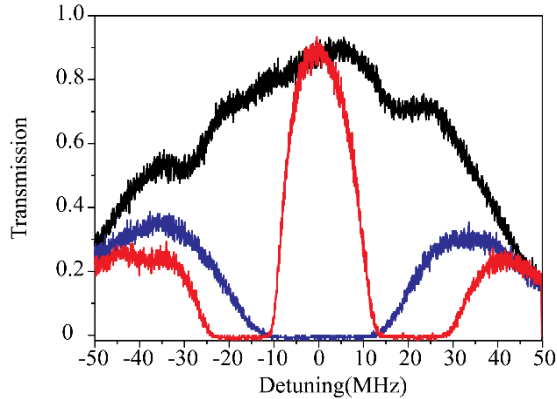

Fig. 8 Transmission spectra (Blue) and EIT spectra (Red). Black line is 50-μs probe light without atomic ensemble in MOT B

**2. The interferometer for storage**

In the Mach-Zehnder interferometer shown in Fig. 1(b), we first used a half-wave plate on the path of vertical polarization after BD40, making light in both paths be the same horizontal polarization. Then a quarter-wave plate was used to make the photon from any of the two paths be $\sigma^+$ polarization before entering the MOT B. After leaving MOT B, another quarter-wave plate reverted their polarization to be horizontal, followed by a half-wave plate in one path making polarization of the photon in that path back to be vertical. This setup was intrinsically stable and without any locking circuit. Under this condition, we could accomplish the storage of an arbitrary state with same efficiency. An attenuator here used was a balance for two terms in Eq. (1).

**3. Error estimations**

Error bars in the experimental data including CHSH and state tomography were estimated from Poisson statistics and using Monte Carlo simulations. Other error bars including the interference curves, $g^{(2)}_{12}$, storage efficiency are from statistical measurement.